\documentclass[prd,draft,showpacs]{revtex4}
\begin{document}
\title{Junction conditions of cosmological perturbations
}
\author{Kenji Tomita }
\affiliation{Yukawa Institute for Theoretical Physics, 
Kyoto University, Kyoto 606-8502, Japan}
\date{\today}

\begin{abstract}
The behavior of perturbations is studied in cosmological models which
consist of two different homogeneous regions connected in a spherical 
shell boundary. The junction conditions for the metric perturbations 
and the displacements of the shell boundary are analyzed and the 
surface densities of the
perturbed energy and momentum in the shell are derived, using 
Mukohyama's gauge-invariant formalism and the Israel
discontinuity condition. In both homogeneous regions, the perturbations 
of scalar, vector and tensor types are expanded using the 3-dimensional 
harmonic functions, but the mode coupling among them is caused in the 
shell by the inhomogeneity. By treating the perturbations with odd
and even parities separately, 
it is found, however, that we can have consistent displacements and
surface densities for given metric perturbations.

\end{abstract}
\maketitle


\section{Introduction}
\label{sec:level1}

In the real Universe, it is known at present that there are many 
superclusters and void-like objects with nonlinear overdense and 
underdense regions which are
distributed everywhere on various scales. Such objects may have
complicated structures in general, but their evolution has often been 
studied in comparatively large-scale cases, using a
simplified assumption of spherical symmetry. 

As general-relativistic examples of spherical treatments we have 
the analysis due to
the Tolman-Bondi-Lemaitre inhomogeneous dust model \cite{tolm,LL},
 \ a single-shell model consisting of two different
homogeneous regions connected with a spherical shell \cite{sakai,bere}, 
\ a self-similar model being described using the self-similar 
solutions \cite{carr,cahill}, \ a model consisting of two homogeneous 
regions connected with the self-similar intermediate region 
\cite{tom95,tom96,tomgrg}, and so on. 

In order to study the dynamical behavior of the large-scale superclusters 
and void-like objects and
their influence on the CMB anisotropy, moreover, it is necessary to
consider the gravitational instability of these models. The general
gauge-invariant formalism for linear perturbations in spherical symmetric 
inhomogeneous models 
was derived by Gerlach and Sengupta \cite{gspert,gsod,gsev,gsharm} in 
four-dimensional spacetimes. Recently the junction of perturbations in 
two homogeneous regions was treated by Mukohyama\cite{muk1,muk2}
 and Kodama et al.\cite{kodama} in higher dimensional spacetimes on 
the basis of the Israel discontinuity condition \cite{israel}.  In this 
paper, we study the behavior of linear 
perturbations in the four-dimensional cosmological models with a 
single shell, using Mukohyama's formalism. Here only one of his doubly 
gauge-invariant variables are used, and in both homogeneous regions
 the perturbations are classified into scalar, vector and tensor 
types in three dimensional space and expanded in terms of three 
dimensional harmonic functions \cite{bard,ks} in order to clarify the 
physical image of perturbations, though he classified 
them in a space of the shell surface. Among them the mode
coupling is caused by the inhomogeneity in the shell. 
By treating the perturbations with odd and even parities separately,
 it is found however that the consistent expressions of the shell 
displacement and the energy-momentum tensor in the shell can be 
derived for given metric perturbations in both regions.   

In \S II, we describe briefly the single-shell model as the background 
model, in which the velocity of the shell is determined through the 
junction condition.  In \S III, we show the basic equations for 
gauge-invariant perturbations and the junction condition based on 
Mukohyama's formalism \cite{muk1,muk2}.
In \S IV, we derive the intrinsic perturbations of metric and extrinsic 
curvature in the shell, and show the displacements of the shell and the 
intrinsic energy-momentum tensor in the shell which are derived from the 
junction conditions, by treating the perturbations with odd-parity and
even-parity separately and imposing a localization condition on the 
wave-number dependence of the shell displacement.   \S V is given to 
the concluding remarks. In Appendix A, the harmonic functions in a 
homogeneous background 
model are shown, which are necessary for the analysis of the description 
of junction conditions. In Appendices B ad C, auxiliary quantities 
$J_{\lambda \mu \nu}$ and $\Theta_{\mu \nu}$ for the perturbed 
gravitational field and the intrinsic energy-momentum tensor in the 
shell, respectively,  are shown.

\section{Background models and the junction conditions}
\label{sec:level2}

The background universe is assumed to consist of two spatially 
homogeneous regions connected by a spherical shell. 
The four-dimensional line-elements in the inner region V$^-$ and the 
outer region V$^+$ are described as
\begin{equation}
  \label{eq:m1}
 ds^2 = g^\pm_{\mu\nu} (dx^\mu)_\pm (dx^\nu)_\pm = [a_\pm (\eta_\pm)]^2 
[- (d \eta_\pm)^2 + dl_\pm^2],
\end{equation}
\begin{equation}
dl_\pm^2 = \gamma^\pm_{ij} (dx^i)_\pm (dx^j)_\pm =
(d \chi_\pm)^2 + [\sigma_\pm (\chi_\pm)]^2 d\Omega^2 ,
\end{equation}
where the suffices $\mu, \nu$ and $i, j$ run from $0$ to $3$ and from $1$ 
to $3$, respectively, \ $x^1_\pm = \chi_\pm$, \  $\sigma_\pm (\chi_\pm) 
= \sin
\chi_\pm, \chi_\pm$ and $\sinh \chi_\pm$ for the spatial curvature 
$K_\pm = 1, 0, -1$, respectively,
and $d\Omega^2 = d\theta^2 + \sin^2 \theta d \varphi^2$.
The shell is a time-like hypersurface $\Sigma$ given as 
\begin{equation}
  \label{eq:m2}
x_\pm^\mu = Z_\pm^\mu (y^M),
\end{equation}
where $y^{M} \ (M = 0, 2, 3)$ denote the intrinsic coordinates in the shell. 
In the following we specify the coordinates in the form   
\begin{equation}
  \label{eq:m3}
x_\pm^A = Z_\pm^A (y^0), \quad x_\pm^a = y^a,
\end{equation}
where the suffices $A$ and $a$ take the values (0, 1) and (2, 3), 
respectively.
When we define the tangential triads \ $e_{(M)\pm}^\mu \ (M = 0, 2, 3)$ by 
\ $e_{(M)\pm}^\mu = \partial x_\pm^\mu/\partial y^M$, the intrinsic metric 
is given by
\begin{equation}
  \label{eq:m4}
q^\pm_{MN} = e^\mu_{(M)\pm} e^\nu_{(N)\pm} g^\pm_{\mu\nu}.
\end{equation}
If we specify $y^0$ as the proper time $\tau$ in the shell, the components 
of the triads can be expressed as
\begin{equation}
  \label{eq:m5}
e^\mu_{(0)\pm} = (\partial x^0/\partial y^0, \partial x^1/\partial y^0, 0, 
0) = F_\pm (1, v_\pm, 0, 0),
\end{equation}
$e^\mu_{(2)\pm} = (0, 0, 1, 0)$ and $e^\mu_{(3)\pm} = (0, 0, 0, 1)$, where 
$F_\pm \equiv \gamma_\pm/a_\pm, \ v_\pm \equiv dx^1_\pm/d x^0_\pm \ 
(= dx^1_\pm/d\eta_\pm)$ and $\gamma_\pm 
\equiv 1/\sqrt{1 - (v_\pm)^2}$. Here we take the units of the light velocity
 $c = 1$. The unit normal vector $n^\mu_\pm$ of $\Sigma$ is
defined by the conditions $n_{\mu \pm} e^\mu_{(M)\pm} = 0$ ($M = 0, 2,3)$ 
and 
$n_{\mu\pm} n^\mu_\pm = 1$, and the components can be expressed as
\begin{equation}
  \label{eq:m6}
n^\mu_\pm = (-v_\pm, 1, 0, 0) \gamma_\pm/a_\pm.
\end{equation}
The line-element in the shell in terms of the intrinsic coordinates is
\begin{equation}
  \label{eq:m7}
ds^2 = q_{MN} dy^M dy^N = - d\tau^2 + R^2 d \Omega^2,
\end{equation}
where $y^0 = \tau, y^2 = \theta, y^3 =
\varphi$, and $R \equiv a_+ \sigma (\chi_+) = a_- \sigma (\chi_-)$.

Since the intrinsic geometry is regular, the induced metric should be 
continuous and we have 
\begin{equation}
  \label{eq:m8}
q_{MN+} = q_{MN-} \equiv q_{MN}.
\end{equation}
The extrinsic curvature of $\Sigma$ is defined by
\begin{equation}
  \label{eq:m9}
K_{MN\pm} = {1 \over 2} e^\mu_{(M)\pm} e^\nu_{(N)\pm} {\cal{L}}_n g_{\mu\nu}
= e^\mu_{(M)\pm} e^\nu_{(N)\pm} [n_{\mu;\nu} + n_{\nu;\mu}], 
\end{equation}
where $\cal{L}$ represents the Lie derivative, $n$ denotes the unit normal of 
$\Sigma$, and a suffix $;\mu$ denotes the four-dimensional covariant derivative 
with respect to $x^\mu$.
 For the energy-momentum tensor $T^{\mu\nu}$, the corresponding surface
component $S_{\mu\nu}$ in the shell is defined using the Gaussian normal 
coordinates as
\begin{equation}
  \label{eq:m10}
S_{\mu\nu} = \lim_{\epsilon \rightarrow 0} \int^\epsilon_{-\epsilon} T_{\mu\nu} 
d\zeta,
\end{equation}
where $\zeta$ is the radial coordinate and $\zeta = 0$ is for the shell 
surface. The expression $S_{MN}$ in the intrinsic coordinates is given by
\begin{equation}
  \label{eq:m11}
S_{MN} = e^\mu_{(M)} e^\nu_{(N)} S_{\mu\nu},
\end{equation}
where $e^\mu_{(M)}$ is the triad in the gaussian normal coordinates.

From the junction conditions derived by Israel, we obtain the relations 
between the jump 
of the extrinsic curvature and the surface energy-momentum tensor $S_{MN}$ as
\begin{equation}
  \label{eq:m12}
[K_{MN}]^\pm = - \kappa^2 (S_{MN} - {1 \over 2} S \ q_{MN}),
\end{equation}
together with the additional relations
\begin{equation}
  \label{eq:m13}
- S^N_{M|N} = [T^n_M]^\pm
\end{equation}
and
\begin{equation}
  \label{eq:m14}
{1 \over 2} (K^M_{N+} + K^M_{N-}) \ S^N_M = [T^n_n]^\pm, 
\end{equation}
where $[\Phi]^\pm \equiv \Phi^+ - \Phi^-$, \ $S \equiv q^{MN} S_{MN},
\ q^{MN}$ 
is the inverse of the induced metric $q_{MN}, \ \kappa^2 \equiv 8\pi G$, 
and
\begin{equation}
  \label{eq:m15}
T^n_{M\pm} = n_{\mu\pm} e^\nu_{(M)\pm} (T^\mu_\nu)_\pm \quad 
T^n_{n\pm} = n_{\mu\pm} n^\nu_\pm (T^\mu_\nu)_\pm,
\end{equation}
\begin{equation}
  \label{eq:m16}
S^N_M = q^{NL} S_{LM}.
\end{equation} 
In the case of dust matter, we have
\begin{equation}
  \label{eq:m17}
T_{\mu\nu\pm} = \rho_\pm u_{\mu\pm} u_{\nu\pm}, \quad S_{MN} = 
\tilde{\sigma} v_{M\pm} v_{N\pm},
\end{equation} 
where $\tilde{\sigma}$ is the surface density of dust matter.
For the metric (\ref{eq:m1}), we have
\begin{equation}
  \label{eq:m18}
K^0_{0\pm} = [\gamma_\pm^2 \dot{v}_\pm + (\dot{a}_\pm/a_\pm) v_\pm]
\gamma_\pm/a_\pm, 
\end{equation} 
and
\begin{equation}
  \label{eq:m19}
K^2_{2\pm} = K^3_{3\pm} = [\sigma'/\sigma + (\dot{a}_\pm/a_\pm) v_\pm]
\gamma_\pm/a_\pm,
\end{equation} 
where $K^M_{N\pm} = q^{ML} K_{LN\pm}$, and a dot and a prime denote
 $\partial/\partial \eta_\pm$ and $\partial/\partial \chi_\pm$, respectively.

By substituting these equations (\ref{eq:m18}) and (\ref{eq:m19}) into the
above Eqs.(\ref{eq:m12}) - (\ref{eq:m14}), we can obtain equations for 
$v_\pm$ and $\tilde{\sigma}$. By solving these equations, 
Sakai et al.\cite{sakai} derived their time evolution and showed that 
$|v_\pm|^2 \approx 10^{-5} \ll 1$. 
Moreover the following constraint equation for $\tilde{\sigma}$ is obtained:
\begin{equation}
  \label{eq:m20}
[\gamma (\sigma' + v H R)]^\pm = -{1 \over 2} \kappa^2 R \tilde{\sigma},
\end{equation} 
where $R \equiv a_+ \sigma_+ = a_- \sigma_-$. 

In the following the suffice $\pm$ is omitted except for the case when 
it is necessary.
Here let us show the relations necessary for the later calculations:
\begin{equation}
  \label{eq:m21}
q_{00} = -1, \quad q_{22} = q_{33}/ \sin^2 \theta = (a\sigma)^2, 
\end{equation} 
\begin{equation}
  \label{eq:m22}
q^{00} = -1, \quad q^{22} = q^{33} \sin^2 \theta = (a\sigma)^{-2}, 
\end{equation} 
\begin{equation}
  \label{eq:m23}
K_{00} = -(\gamma/a)[\gamma^2 \dot{v} +(\dot{a}/a)v], 
\quad K_{22} = K_{33}/ \sin^2 \theta = a\gamma\sigma^2 [\sigma'/\sigma + 
(\dot{a}/a)v], 
\end{equation} 
where $K \equiv q^{MN}K_{MN} = (\gamma/a)[2\sigma'/\sigma +3(\dot{a}/a)v
+ \gamma^2 \dot{v}]$ and $K^{00} = K_{00}, K^{22} = K^{33} \sin^2 \theta =
K_{22}/(a\sigma)^{4}$.
\begin{eqnarray}
  \label{eq:m24}
K_{00}-Kq_{00} &=& 2(\gamma/a)[\sigma'/\sigma +(\dot{a}/a)v], \cr
K_{22}-Kq_{22} &=& [K_{33}-Kq_{33}]/ \sin^2 \theta = 
-a\gamma\sigma^2 [\sigma'/\sigma + 2(\dot{a}/a)v + \gamma^2 \dot{v}]. 
\end{eqnarray}

\section{Perturbations and the junction conditions}
\label{sec:level3}
The perturbations of $g_{\mu\nu}$ and $Z^\mu$ around the background 
$(\bar{g}_{\mu\nu}$ and $\bar{Z}^\mu$) are considered:
\begin{equation}
  \label{eq:b1}
g_{\mu\nu} = \bar{g}_{\mu\nu} + \delta g_{\mu\nu}, \quad Z^\mu =
\bar{Z}^\mu + \delta Z^\mu.
\end{equation}
The perturbed tangent vectors are expressed as
\begin{equation}
  \label{eq:b2}
e^\mu_{(M)} = \bar{e}^\mu_{(M)} - {\cal{L}}_{\delta Z} \bar{e}^\mu_{(M)}. 
\end{equation}
For the perturbations of the induced metric:
\begin{equation}
  \label{eq:b3}
q_{MN} = \bar{q}_{MN} + \delta q_{MN},
\end{equation}
Mukohyama derived the expression
\begin{equation}
  \label{eq:b4}
\delta q_{MN} = e^\mu_{(M)} e^\nu_{(N)} (\delta g_{\mu\nu} + 
{\cal{L}}_{\delta Z} \bar{g}_{\mu\nu}). 
\end{equation}
The perturbed normal is expressed as
\begin{equation}
  \label{eq:b5}
\delta n^\mu = {1 \over 2} \bar{n}^\mu \bar{n}^\nu \bar{n}^\lambda \delta 
g_{\nu\lambda} + e^\mu_{(M)} q^{MN} \bar{n}_\nu {\cal{L}}_{\delta Z} 
e^\nu_{(N)}
\end{equation}
and 
\begin{equation}
  \label{eq:b6}
\delta K_{MN} = {1 \over 2} e^\mu_{(M)} e^\nu_{(N)} [{\cal{L}}_{\delta Z} 
\bar{g}_{\mu\nu} + {\cal{L}}_{\delta Z} {\cal{L}}_n \bar{g}_{\mu\nu}
-2 n^\lambda \delta \Gamma_{\lambda\mu\nu}],
\end{equation}
where 
\begin{equation}
  \label{eq:b7}
\delta \Gamma_{\lambda\mu\nu} \equiv {1 \over 2}(\delta g_{\lambda\mu;\nu}
+ \delta g_{\lambda\nu;\mu} - \delta g_{\mu\nu;\lambda}). 
\end{equation}
This perturbed extrinsic curvatures are reduced to
\begin{equation}
  \label{eq:b8}
\delta K_{MN} = {1 \over 2} I K_{MN} - {1 \over 2} n^\lambda e^\mu_{(M)}
e^\nu_{(N)} J_{\lambda\mu\nu},
\end{equation}
where 
\begin{equation}
  \label{eq:b9}
I \equiv n^\mu n^\nu (\delta g_{\mu\nu} + 2 \delta Z_{\mu;\nu})
\end{equation}
and 
\begin{equation}
  \label{eq:b10}
J_{\lambda\mu\nu} \equiv 2 \delta \Gamma_{\lambda\mu\nu} + \delta 
Z_{\lambda;\mu\nu} + \delta Z_{\lambda;\nu\mu} + (R_{\alpha\mu\lambda\nu}
+ R_{\alpha\nu\lambda\mu}) \delta Z^\alpha. 
\end{equation}
Under the gauge transformation
\begin{equation}
  \label{eq:b11}
x^\mu \rightarrow {x'}^\mu = x^\mu + \xi^\mu,
\end{equation}
$\delta_{\mu\nu}$ and $\delta Z^\mu$ transform as
\begin{equation}
  \label{eq:b12}
\delta g_{\mu\nu} \rightarrow \delta g_{\mu\nu} - \xi_{\mu;\nu} 
- \xi_{\nu;\mu}, \quad \delta Z^\mu \rightarrow \delta Z^\mu + \xi^\mu,
\end{equation}
and it is found that $\delta q_{MN}$ and $\delta K_{MN}$ are invariant under this gauge transformation. The junction conditions for the perturbations of
metric and extrinsic curvature which was shown by Mukohyama\cite{muk2} 
are expressed as
\begin{equation}
  \label{eq:b13}
\delta q_{MN+} = \delta q_{MN-} 
\end{equation}
and 
\begin{equation}
  \label{eq:b14}
[\delta \tilde{K}_{MN}]^\pm = - \kappa^2 (\delta \tilde{S}_{MN}
 - {1 \over 2} \delta \tilde{S}\ q_{MN}),
\end{equation}
where 
\begin{equation}
  \label{eq:b15}
\delta \tilde{K}_{MN} \equiv \delta K_{MN} - {1 \over 2} (K^L_M \delta q_{LN}
+ K^L_N \delta q_{LM}),
\end{equation}
and
\begin{equation}
  \label{eq:b16}
\delta \tilde{S}_{MN} \equiv \delta S_{MN} - {1 \over 2} (S^L_M \delta q_{LN}
+ S^L_N \delta q_{LM})
\end{equation}
with $S = q^{MN} S_{MN}$ and $\delta \tilde{S} = q^{MN} \delta S_{MN}$.
The equation (\ref{eq:b14}) is consistent with a relation
\begin{equation}
  \label{eq:b17}
[\delta \tilde{K}_{MN}]^\pm = -\kappa^2 \{\delta S_{MN} - 
{1 \over 2} (\delta S q_{MN} + S \delta q_{MN})\},
\end{equation}
so that
\begin{equation}
  \label{eq:b18}
- \kappa^2 \delta S_{MN} = [\delta K_{MN} - \delta K q_{MN} 
- K \delta q_{MN}]^\pm,
\end{equation}
where $\delta K = \delta q^{MN} K_{MN} + q^{MN} \delta K_{MN} = 
q^{MN} \delta K_{MN} - K^{MN} \delta q_{MN}$.

When the metric perturbations in the two regions V$^+$ and V$^-$ are given,
we can determine $\delta q_{MN}, \delta K_{MN}$ and $\delta S_{MN}$ 
using Eqs.
(\ref{eq:b4}), (\ref{eq:b6}) and (\ref{eq:b18}), and find the conditions 
which are imposed on the shell displacements, using Eq.(\ref{eq:b13}).

\section{Perturbed quantities in the boundary shell}
\label{sec:level4}

In this section we express first the perturbations of the induced metric 
and the extrinsic curvature, using the gauge-invariant variables 
representing  metric 
perturbations and the shell displacement, and derive the perturbations of 
the energy-momentum tensors in the shell.
The metric perturbations and the harmonic functions in the homogeneous 
regions are shown in Appendix A.

\subsection{Metric perturbations}
The metric perturbations in Eq. (\ref{eq:b4}) are rewritten in terms of 
$x^\mu$ as 
\begin{equation}
  \label{eq:c1}
\delta q_{MN} dy^M dy^N = (\delta g_{\mu\nu} + \delta Z_{\mu;\nu} + 
\delta Z_{\nu;\mu}) dx^\mu dx^\nu,
\end{equation}
where $\delta Z_\mu$ represents the displacement of the boundary shell.
$\delta Z_0$ and $\delta Z_i$ are here treated as spatially 
3-dimensional quantities in accord with the metric perturbations.
In principle they are the quantities given in the 
shell. In Mukohyama's formalism\cite{muk2} the displacements of the 
shell were regarded as functions of only intrinsic variables $\tau,
\theta$ and $\varphi$ which are spatially 2-dimensional.  
In order to recover the local nature of $\delta Z_0$ and $\delta Z_i$ in
our case, we impose later a localization condition, under which 
their behavior is constrained so that their values may be nonzero only
in the neighborhood of the shell.
The coefficients in the right-hand side of this equation are expressed
 as follows using the gauge-invariant variables. 

\subsubsection{Scalar perturbations}
\begin{eqnarray}
  \label{eq:c2}
\delta g_{00} &+& 2 \delta Z_{0;0} = \int k^2 dk [-2 a^2 \Phi_A + 2a(\phi_0/a)^.] Q,\cr
\delta g_{0i} &+& \delta Z_{i;0} + \delta Z_{0;i} = \int k^2 dk [a^2(\phi_L/a^2)^. + 
\phi_0] Q_{,i},\cr
\delta g_{ij} &+& \delta Z_{i;j}+ \delta Z_{j;i} = \int k^2 dk [(2a^2 \Phi_H 
- 2{\dot{a} \over a} \phi_0) \gamma_{ij} Q - k \phi_L (Q_{i|j} + Q_{j|i})],
\end{eqnarray}
where the suffix $|i$ denotes the covariant derivative in the 
3-dimensional space with $dl^2 = \gamma_{ij} dx^i dx^j$, \ 
$\delta Z_0$ and $\delta Z_i$ are expanded in terms of harmonic functions
as $\delta Z_0 = \int k^2 dk \ z_0 Q$ and 
$\delta Z_i = \int k^2 dk \ z_L Q_i,$
and \ ${\bf k}$ is the wave-vector, whose length is $k  = |{\bf k}|$.
 Here, $\Phi_A, \Phi_H, \phi_0$ 
and $\phi_L$ are the gauge-invariant variables defined by
\begin{eqnarray}
  \label{eq:c3}
\Phi_A &\equiv& A + {1\over k}\dot{B} + {1\over k}{\dot{a} \over a}B
- {1\over k^2}\left(\ddot{H}_T + {\dot{a} \over a}\dot{H}_T\right),  \cr
\Phi_H &\equiv& H_L + {1 \over 3}H_T+ {1\over k}{\dot{a} \over a}B
- {1\over k^2}{\dot{a} \over a}\dot{H}_T,  \cr
\phi_0 &\equiv& z_0 + (a^2/k)(B - \dot{H}_T/k),\cr
\phi_L &\equiv& -z_L/k + a^2H_T/k^2,
\end{eqnarray}
where various expressions for the metric perturbations and harmonic functions 
are shown in Appendix A. Here we use normalized harmonic functions.

Since $dx^0 = F dy^0, dx^1 = F v dy^0, dx^2 = dy^2$ and $dx^3 = dy^3$
with $F \equiv e^0_{(0)} =\gamma/a$, we obtain
\begin{eqnarray}
  \label{eq:c5}
\delta q_{00} &=& \int k^2 dk \ F^2 [\zeta_0 Q + 2v \zeta_L Q_{,1} + 
v^2(\zeta_{LL} \gamma_{11} Q +2k^2 \phi_L Q_{11})],\cr
\delta q_{0a} &=& \int k^2 dk \ F [\zeta_L Q_{,a} + 2v k^2 \phi_L Q_{1a}],\cr
\delta q_{ab} &=& \int k^2 dk \ [\zeta_{LL} \gamma_{ab} Q + 2k^2 
\phi_L Q_{ab}],
\end{eqnarray}
where $\zeta_0 \equiv -2a^2 \Phi_A + 2a(\phi_0/a)^., \ \zeta_L \equiv
a^2(\phi_L/a^2)^. + \phi_0$ and $\zeta_{LL} \equiv 2a^2 \Phi_H - 
2(\dot{a}/ a)\phi_0 -{2 \over 3}k^2\phi_L$. 
For $I$ defined in Eq.(\ref{eq:b9}), we obtain
\begin{equation}
  \label{eq:c6}
(a/\gamma)^2 I = \int k^2 dk \left\{\left[2a^2\Phi_H -2{\dot{a} \over a}
\phi_0\right]Q + 2\phi_L Q_{,11} + 2v[a^2(\phi_L/a^2)^. +\phi_0] Q_{,1}
+ v^2 [-2a^2 \Phi_A + a(\phi_0/a)^.] Q \right\}.
\end{equation}
Equations (\ref{eq:c5}) are rewritten using spherical harmonics as follows:
\begin{eqnarray}
  \label{eq:c6a}
\delta q_{00} &=& \int k^2 dk F^2 \left\{[\zeta_0 \Pi + 2v \zeta_L \Pi' + 
v^2\left[\zeta_{LL} \Pi +2\phi_L \left(\Pi''+{k^2 \over 3}\Pi \right)\right]
\right\} Y_{lm},\cr
\delta q_{0a} &=& \int k^2 dk F [\zeta_L \Pi + 2v \phi_L (\Pi' - {\sigma' \over \sigma} 
\Pi)] Y_{lm,a},\cr
\delta q_{ab} &=& \int k^2 dk \left[\phi_L \Pi \tilde{Y}_{ab} + 
\left\{\zeta_{LL} + 2\phi_L 
\left[\Pi' {\sigma' \over \sigma} +\left({k^2 \over 3} - {l(l+1) 
\over 2\sigma^2}\right)\Pi \right] \right\}
\gamma_{ab} Y_{lm}\right],
\end{eqnarray}
where $Q = \Pi_l (k,\chi) Y_{lm} (\theta, \varphi)$, the suffix $l$ in 
$\Pi_l$ is omitted, $Y_{lm} (\theta, \varphi)$ is the spherical harmonics 
and $\tilde{Y}_{ab}$ is a 2-dimensional traceless tensor defined by  
\begin{equation}
  \label{eq:c6b}
\tilde{Y}_{ab} \equiv Y_{lm \| ab} + {l(l+1) \over 2\sigma^2}\gamma_{ab} 
Y_{lm}.
\end{equation}
Here  $a$ and $b$ take the value $2$ or $3$ and \ $\|$ denotes the covariant 
derivative with respect to the space with $d\Omega^2 = d\theta^2 + 
\sin^2 \theta  d\varphi^2$.  

In these equations, dots mean $\partial/\partial \eta_+$ and 
$\partial/\partial \eta_-$ in V$^+$ and V$^-$, respectively,  
primes mean $\partial/\partial \chi_+$ and 
$\partial/\partial \chi_-$ in V$^+$ and V$^-$, respectively, 
$\Pi_+$ and $\Pi_-$ are equal to $\Pi(\chi_+)$ and $\Pi(\chi_+)$, and
a suffix $k$ is omitted here and in the following. Here, if we change
the time variables $\eta_+$ and $\eta_-$ to the common variable 
$\tau (= y^0)$ in the shell by $\partial/\partial \eta_\pm = 
(a_\pm/\gamma_\pm) \partial/\partial \tau$, then we have
\begin{eqnarray}
  \label{eq:c6c}   
\zeta_0 &=& -2 a^2 \Phi_A + {2 \over \gamma} a^2 (\phi_0/a)_{,\tau}, \cr
\zeta_L &=& a^3 (\phi_L/a^2)_{,\tau} + \phi_0, \cr
\zeta_{LL} &=& 2a^2 \Phi_H - 2a_{,\tau} \phi_0 - {2 \over 3} k^2 \phi_L.
\end{eqnarray}

\subsubsection{Vector perturbations}
The coefficients of Eq.(\ref{eq:c1}) are
\begin{eqnarray}
  \label{eq:c7}
\delta g_{00} &+& 2 \delta Z_{0;0} = 0,\cr
\delta g_{0i} &+& \delta Z_{i;0} + \delta Z_{0;i} = \int k^2 dk [-a^2\Psi
+ a^2(\phi_T/a^2)^.] V_i,\cr
\delta g_{ij} &+& \delta Z_{i;j}+ \delta Z_{j;i} = \int k^2 dk [-2k \phi_T] V_{ij},
\end{eqnarray}
where $\delta Z_0 = 0$ and $\delta Z_i = \int k^2 dk z_T V_i$, and 
$\Psi$ and $\phi_T$ 
are the gauge-invariant variables defined as follows:
\begin{equation}
  \label{eq:c8}
\Psi \equiv B - {1 \over k} \dot{H}_T^{(1)}, \quad \phi_T \equiv z_T - {1 \over k} 
a^2 \dot{H}_T^{(1)}. 
\end{equation}
The induced metric perturbations are 
\begin{eqnarray}
  \label{eq:c9}
\delta q_{00} &=& -2F^2 v \int k^2 dk \{[-a^2\Psi + a^2(\phi_T/a^2)^.]V_1
 + vk\phi_T V_{11}\},\cr
\delta q_{0a} &=& F \int k^2 dk \{[-a^2\Psi + a^2(\phi_T/a^2)^\cdot]V_a
 -2vk \phi_T V_{1a}\},\cr
\delta q_{ab} &=&  -2\int k^2 dk \ (k \phi_T) V_{ab},
\end{eqnarray}
where $V_{1a} = - {1 \over 2k} [V_{1,a} + R^2(V_a/R^2)_{,1}]$.
For $I$ we obtain
\begin{equation}
  \label{eq:c10}
(a/\gamma)^2 I = -2\int k^2 dk \{v[a^2\Psi +a^2(\phi_T/a^2)^.] V_1 + 
k\phi_T V_{11}\}.
\end{equation}
Equations (\ref{eq:c9}) are rewritten using spherical harmonics in the cases
of odd and even parities as follows:

\medskip
\noindent (2-1) The {\it odd-parity} case

Since $V_1 = 0$ (cf. Appendix A), we have 
\begin{eqnarray}
  \label{eq:c10a}
\delta q_{00} &=& 0, \cr
\delta q_{0a} &=&  F\int k^2 dk \{[-a^2 \Psi + (a^3/\gamma)
(\phi_T/a^2)_{,\tau}]V_a + v \phi_T [V_{a,1} - 2(\sigma'/\sigma)V_a]\}, \cr
\delta q_{ab} &=& \int k^2 dk \phi_T [V_{a \| b} + V_{b \| a}],
\end{eqnarray}
where  $(V_2, V_3) = \sigma(\chi) 
\Pi (-Y_{lm,3}/\sin \theta,  Y_{lm,2} \sin \theta)$. These 3-dimensional
vector perturbations are also 2-dimensional vector perturbations.

\medskip
\noindent (2-2) The {\it even-parity} case

 Harmonic functions are 
\begin{eqnarray}
  \label{eq:c10b}
V_1 &=& l(l+1) (\Pi/\sigma) Y_{lm}, \cr
(V_2, V_3) &=& (\Pi\sigma)' (Y_{lm,2}, Y_{lm,3}),
\end{eqnarray}
and the intrinsic metric perturbations are
\begin{eqnarray}
  \label{eq:c10c}
\delta q_{00} &=& -2F^2 v \int k^2 dk \left[-a^2\Psi + (a^3/\gamma)
(\phi_T/a^2)_{,\tau}
-v \phi_T\left({\Pi' \over \Pi}-{\sigma' \over \sigma}\right)\right] V_1, \cr
\delta q_{0a} &=&  F\int k^2 dk \left\{[-a^2\Psi + (a^3/\gamma)
(\phi_T/a^2)_{,\tau} ]V_a +
v\phi_T \left[V_{a,1} - 2(\sigma' / \sigma)V_a + V_{1,a}\right]\right\}, \cr
\delta q_{ab} &=& \int k^2 dk \phi_T [V_{a\| b} + V_{b\| a} + 2 \gamma_{ab} 
{\sigma' \over \sigma}V_1].
\end{eqnarray}
These expressions are reduced  to
\begin{eqnarray}
  \label{eq:c10d}
\delta q_{00} &=& -2F^2 v \int k^2 dk \left[-a^2\Psi + (a^3/\gamma)
(\phi_T/a^2)_{,\tau}
-v \phi_T\left({\Pi' \over \Pi}-{\sigma' \over \sigma}\right)\right] l(l+1) 
(\Pi/\sigma) Y_{lm}, \cr
\delta q_{0a} &=& F \int k^2 dk \{[-a^2\Psi + (a^3/\gamma)(\phi_T/a^2)_{,\tau}
 ](\Pi\sigma)' +
v\phi_T [(\Pi\sigma)'' - 2(\sigma'/ \sigma)(\Pi\sigma)'
+ l(l+1)(\Pi/\sigma)]\} Y_{lm,a}, \cr
\delta q_{ab} &=& \int k^2 dk \phi_T \{2(\Pi\sigma)' \tilde{Y}_{ab} + {l(l+1) \over 
\sigma^2}[2\Pi\sigma' - (\Pi\sigma)'] \} \gamma_{ab} Y_{lm}.
\end{eqnarray}
In this form the dependence on spherical harmonics is found to be quite 
the same as that of scalar perturbations.

\subsubsection{Tensor perturbations}

The non-zero coefficient of Eq.(\ref{eq:c1}) is
\begin{equation}
  \label{eq:c11}
\delta g_{ij} + \delta Z_{i;j}+ \delta Z_{j;i} = -2 a^2 \int k^2 dk H_T^{(2)} 
G_{ij},
\end{equation}
and the induced metric components are
\begin{eqnarray}
  \label{eq:c12}
\delta q_{00} &=& -2a^2 F^2 v^2 \int k^2 dk H_T^{(2)} G_{11},\cr
\delta q_{0a} &=& -2a^2 F v \int k^2 dk H_T^{(2)} G_{1a},\cr
\delta q_{ab} &=& -2a^2 \int k^2 dk H_T^{(2)} G_{ab}.
\end{eqnarray}
For $I$ we obtain
\begin{equation}
  \label{eq:c13}
(a/\gamma)^2 I = -2a^2 \int k^2 dk H_T^{(2)} G_{11}.
\end{equation}
Equations (\ref{eq:c12}) are rewritten using spherical harmonics in the cases
of odd and even parities as follows:

\medskip
\noindent (3-1) The {\it odd-parity} case

If we define a 2-dimensional axial vector $W_{a}$ by
\begin{equation}
  \label{eq:c13a}
W_{a} \equiv (- Y_{lm,3}/\sin \theta, Y_{lm,2} \sin \theta)
\end{equation}
for $a = 2, 3$, we obtain 
\begin{eqnarray}
  \label{eq:c13b}
\delta q_{00} &=& 0, \cr
\delta q_{0a} &=& F v \int k^2 dk H_T^{(2)} (l-1)(l+2) \Pi W_a,\cr
\delta q_{ab} &=& \int k^2 dk H_T^{(2)} (\Pi \sigma^2)' (W_{a \| b} + W_{b \| a}).
\end{eqnarray}
When we compare $W_a$ with $V_a$ in the odd-parity case of vector 
perturbations, we have the relation $V_a = (\Pi \sigma) W_a$, so that 
with respect to the two-dimensional angular dependence the tensor 
perturbations in the odd-parity case have the same form as
vector perturbations in the odd-parity case.   

\medskip
\noindent (3-2) The {\it even-parity} case
\begin{eqnarray}
  \label{eq:c13c}
\delta q_{00} &=& (Fv)^2 \int k^2 dk H_T^{(2)} L(\Pi/\sigma^2) Y_{lm}, \cr
\delta q_{0a} &=& F v \int k^2 dk H_T^{(2)} (l-1)(l+2){(\Pi \sigma)' \over \sigma}
Y_{lm,a},\cr
\delta q_{ab} &=& \int k^2 dk H_T^{(2)} \left[2{\cal G}^n_l \tilde{Y}_{ab} 
- {L \over 2} \Pi\gamma_{ab} Y_{lm}\right], 
\end{eqnarray}
where the definitions of $L$ and ${\cal G}^n_l$ are given in Appendix A. 
In this form the dependence on spherical harmonics is found to be quite 
the same as that of scalar perturbations, as well as that of vector 
perturbations in the even-parity case.

\subsection{Perturbations of the extrinsic curvature and the 
energy-momentum tensor \\
in the shell}
Now let us calculate the perturbed extrinsic curvature $\delta K_{MN}$ 
using Eq.(\ref{eq:b8}) and derive the perturbed energy-momentum tensor 
$\delta S_{MN}$ from Eq.(\ref{eq:b18}). The former expression is rewritten as
\begin{equation}
  \label{eq:c14}
\delta K_{MN} = {1 \over 2} I \ K_{MN} - {1 \over 2} {\gamma \over a} 
\int k^2 dk {\cal{F}}_{MN},
\end{equation}
where ${\cal{F}}_{MN}$ is defined by ${\cal{F}}_{MN} \equiv n^\lambda 
e^\mu_{(M)} 
e^\nu_{(N)} J_{\lambda\mu\nu}$. They have the following components:
\begin{eqnarray}
  \label{eq:c15}
{\cal{F}}_{00} &=& (\gamma/a)^2[J_{100} + v(J_{000}+2 J_{101}) +
v^2 (J_{111} +2J_{001}) + v^3J_{011}],\cr
{\cal{F}}_{22} &=& J_{122} + v J_{022},\cr
{\cal{F}}_{33} &=& J_{133} + v J_{033},\cr
{\cal{F}}_{23} &=& J_{123} + v J_{023},\cr
{\cal{F}}_{02} &=& (\gamma/a)[J_{102} +v(J_{002}+J_{112}) +v^2 J_{012}],\cr
{\cal{F}}_{03} &=& (\gamma/a)[J_{103} +v(J_{003}+J_{113}) +v^2 J_{013}],
\end{eqnarray}  
where ${\cal{F}}_{20} = {\cal{F}}_{02}, {\cal{F}}_{30} = {\cal{F}}_{03},$ and 
${\cal{F}}_{32} = {\cal{F}}_{23}$. The expressions of $J_{\lambda\mu\nu}$ in
terms of gauge-invariant variables of metric perturbations and displacements
of the boundary shell are shown in Appendix B.

The energy-momentum tensor in the shell is expressed as
\begin{eqnarray}
  \label{eq:c16}
-\kappa^2 \delta S_{MN} &=& [\tilde{\Theta}_{MN}]^\pm,\cr
\tilde{\Theta}_{MN} &\equiv& \int k^2 dk \Theta_{MN} \equiv 
{1 \over 2} I (K_{MN} - K q_{MN}) - {1 \over 2}
{\gamma \over a}\int k^2 dk \tilde{\cal{F}}_{MN} - K \delta q_{MN},
\end{eqnarray}
where 
\begin{equation}
  \label{eq:c17}
\tilde{{\cal{F}}}_{MN} \equiv {\cal{F}}_{MN} - q^{KL}{\cal{F}}_{KL}\ q_{MN}
\end{equation}
and their components are
\begin{eqnarray}
  \label{eq:c18}
\tilde{\cal{F}}_{00} &=& {1 \over (a\sigma)^2}[J_{122} + vJ_{022} + 
(J_{133}+ vJ_{033})/\sin^2\theta],\cr
\tilde{\cal{F}}_{22} &=& (\sigma\gamma)^2 [J_{100} + v(J_{000} +2J_{101})
+v^2(J_{111}+2J_{001}) +v^3J_{011}] - (J_{133}+ vJ_{033})/\sin^2\theta,\cr
\tilde{\cal{F}}_{33} &=& (\sigma\gamma)^2 \sin^2\theta [J_{100} + v(J_{000} 
+2J_{101})+v^2(J_{111}+2J_{001}) +v^3J_{011}]\cr
 && - (J_{122}+ vJ_{022}) \sin^2\theta.
\end{eqnarray}
The components of $\Theta_{MN}$ in terms of the gauge-invariant variables 
are obtained through lengthy calculations and are shown in Appendix C
in the expanded form with respect to the shell velocity $v$.

\subsection{Junction of perturbations}
From now let us discuss the junction of metric perturbations given by 
$[\delta q_{MN}]^\pm = 0$ in Eq.(\ref{eq:b13}), in order to analyze
 the displacements of the shell. If the displacements are obtained 
for given metric
perturbations in both regions, we can determine the perturbed 
energy-momentum tensor in the surface ($\delta S_{MN}$)
using the above equations (\ref{eq:c16}).

  The following analyses in the shell are performed separately 
in the odd-parity and even-parity cases of metric perturbations,
and all perturbations in each case are treated together in spite of 
the difference in their 3-dimensional behaviors. 
This is because the 2-dimensional angular dependence of metric 
perturbations in each case is the same and indistinguishable, as was 
shown in the previous subsection. 

\subsubsection{Odd-parity perturbations}
Here we consider vector and tensor perturbations together, and
from the condition $[(\delta q_{MN})_{\rm (vector)} + 
(\delta q_{MN})_{\rm (tensor)}]^\pm = 0$, we obtain the following two
independent equations:
\begin{eqnarray}
  \label{eq:c19}
&&\left[\int k^2 dk \left\{[-\gamma a \Psi + a^2(\phi_T/a^2)_{,\tau}](\Pi 
\sigma) + {\gamma \over a}v [\phi_T (\Pi'\sigma -\Pi\sigma') + 
(l-1)(l+2) H_T^{(2)}\Pi]\right\}\right]^\pm  = 0, \cr
&&\left[ \int k^2 dk \left\{\phi_T \sigma\Pi + H_T^{(2)}(\sigma^2 \Pi)'
\right\} \right]^\pm = 0,
\end{eqnarray}
where the $x^0 (= \eta)$ derivative was changed to the $\tau$ derivative by
$\partial/\partial \eta = (a/\gamma)\partial/\partial \tau$.

The metric perturbations $\Psi$ and $H^{(2)}_T$ are given as functions of 
$\tau$ and $k$ in both regions. In order to recover the original local 
nature of $\phi_T$, on the other hand, we impose the following 
localization condition on it using a function $d (\chi)$:
\begin{equation}
  \label{eq:c19a}
\int \phi_{T\pm}\Pi_\pm \sigma_\pm k^2 dk = \bar{\phi}_{T\pm} (\tau) 
d_\pm (\chi),
\end{equation} 
respectively, where $\bar{\phi}_{T\pm} (\tau)$ are functions of only 
$\tau$ and the functions $d_\pm (\chi)$ corresponding to the shell 
boundary ($\chi = \chi_b$) and the small width $\epsilon$ are defined as
\begin{equation}
  \label{eq:c19b}
d_+(\chi) = 1 \quad {\rm for}\quad \chi_b \leq \chi \leq \chi_b + \epsilon 
\quad {\rm and}\quad 0 \quad {\rm for}\quad 0  \leq \chi <\chi_b 
\quad {\rm or} \quad \chi >\chi_b + \epsilon, 
\end{equation} 
and
\begin{equation}
  \label{eq:c19c}
d_-(\chi) = 1 \quad {\rm for}\quad \chi_b - \epsilon \leq \chi \leq \chi_b 
\quad {\rm and} \quad 0 \quad {\rm for} \quad 0 \leq \chi <\chi_b - 
\epsilon \quad {\rm or} \quad \chi >\chi_b.
\end{equation} 
The above condition means that the total displacement ($\int \phi_{T\pm}
\Pi_\pm \sigma_\pm k^2 dk$) has the nonzero value only in the 
neighborhood of the shell. 
Here we assume that $\epsilon$ is small enough compared with $\chi_b$.
Then we find that $\phi_{T\pm}$ are inversely expressed as
\begin{equation}
  \label{eq:c19d}
\phi_{T\pm} = \bar{\phi}_{T\pm}(\tau) {1 \over k^2} \int^\infty_0 \Pi_\pm
(\tilde{\chi})\sigma_\pm (\tilde{\chi}) d_\pm (\tilde{\chi}) d\tilde{\chi} \ \simeq \ \bar{\phi}_{T\pm}(\tau) \epsilon 
k^{-2} \Pi_\pm (k,\chi_b)
\end{equation} 
in terms of $\bar{\phi}_{T\pm}(\tau)$, using the orthonormal 
relations(\cite{wilson,harr}) of $\Pi$ which are expressed as
\begin{eqnarray}
  \label{eq:c19e}
\int^\infty_0 d \chi  \Pi_\pm (k,\chi)\Pi_\pm (\tilde{k},\chi) \sigma^2(\chi) &=&
\delta (k - \tilde{k}),  \cr
\int^\infty_0 dk \Pi_\pm (k,\chi)\Pi_\pm (k,\tilde{\chi}) \sigma(\chi) 
\sigma(\tilde{\chi})
&=& \delta (\chi - \tilde{\chi})
\end{eqnarray}
for $\Pi_\pm$ with common $l$. Here it should be noticed that the 
orthonormal relations require the integration interval $[0, \infty]$ 
of $\chi$, but it is formal and independent of the physical interval of 
$\chi$ in the two homogeneous regions.
Eq.(\ref{eq:c19d}) shows the $k$ dependence of $\phi_{T\pm}$.
Under the above condition, the two unknown variables $\bar{\phi}_{T+}$ 
and $\bar{\phi}_{T-}$ are determined by solving the above
two equations, as follows.

First we obtain from the latter equation of Eq.(\ref{eq:c19})
\begin{equation}
  \label{eq:c20}
\bar{\phi}_{T-} = \bar{\phi}_{T+} + \int k^2 dk [H_T^{(2)} 
(\sigma^2 \Pi)']^\pm.
\end{equation} 
Next let us differentiate the latter equation 
with respect to $\tau$. Then we obtain
\begin{equation}
  \label{eq:c21}
\left[\int k^2 dk \left\{\phi_{T,\tau} (\Pi\sigma) + H_{T,\tau}^{(2)} (\Pi\sigma^2)' + 
[\phi_T (\Pi\sigma)' + H_T^{(2)}(\Pi \sigma^2)'']\gamma v/a \right\}\right]^\pm = 0.
\end{equation} 
Eliminating $\phi_{T,\tau}$ from the former of Eq.(\ref{eq:c19}) and
Eq.(\ref{eq:c21}), we obtain
\begin{equation}
  \label{eq:c22}
\left[\int k^2 dk \left\{2 \phi_T \left[{a_\tau \over a}\sigma + \gamma 
(v/a)\sigma')\right]\Pi +\gamma a \Psi \Pi \sigma + H_{T,\tau}^{(2)} 
(\Pi\sigma^2)'+ \gamma (v/a) H_T^{(2)}[(\Pi \sigma^2)'' 
-(l-1)(l+2)\Pi]\right\}\right]^\pm = 0.
\end{equation} 
Substituting Eqs.(\ref{eq:c19a}) and (\ref{eq:c19d}) to this equation, 
moreover, we obtain
\begin{equation}
  \label{eq:c22a}
\left[2 \bar{\phi}_T \left({a_\tau \over a}\sigma + \gamma 
(v/a)\cal{L}\right)\right]^\pm + \left[\int k^2 dk 
\left\{\gamma a \Psi \Pi\sigma + H_{T,\tau}^{(2)} 
(\Pi\sigma^2)'+ \gamma (v/a) H_T^{(2)}[(\Pi \sigma^2)'' 
-(l-1)(l+2)\Pi]\right\}\right]^\pm = 0,
\end{equation} 
where
\begin{equation}
  \label{eq:c22b}
{\cal{L}} \equiv \int dk \int d\tilde{\chi} d(\tilde{\chi}) \Pi(\tilde{\chi})
\sigma(\tilde{\chi})\ \Pi(\chi) \sigma'(\chi). 
\end{equation}
In Appendix D, ${\cal{L}}$ is rewritten and reduced to a compact form.
Eliminating $\bar{\phi}_{T-}$ using Eq.(\ref{eq:c20}), therefore, 
we obtain 
\begin{eqnarray}
  \label{eq:c23}
2 \bar{\phi}_{T+}  \Big[{a_\tau \over a} &+& {\gamma v \over a}{\cal{L}}
\Big]^\pm = \left({a_\tau \over a} + {\gamma v \over a}{\cal{L}} \right)_- 
\left[\int k^2 dk H_T^{(2)} (\Pi\sigma^2)'\right]^\pm \cr
&-& \left[\int k^2 dk \left\{\gamma a \Psi \Pi\sigma + 
H_{T,\tau}^{(2)} (\Pi\sigma^2)'+ \gamma (v/a) H_T^{(2)}[(\Pi 
\sigma^2)'' -(l-1)(l+2)\Pi]\right\}\right]^\pm.
\end{eqnarray}
Thus Eqs.(\ref{eq:c20}) and (\ref{eq:c23}) give us $\bar{\phi}_{T+}$ and 
$\bar{\phi}_{T-}$  in terms of given 3-dimensional metric perturbations.
Here let us assume that $v$ vanishes for simplicity. This condition holds
in a good approximation, because $v \ (\approx 10^{-5})$ is very small 
in our background models, as was shown by Sakai et al. (cf. \S 2). 
Then $\bar{\phi}_T$ is reduced to
\begin{equation}
  \label{eq:c24}
\bar{\phi}_{T+} = -{1 \over 2} \left[{a_\tau \over a}\right]_\pm^{-1} 
\left\{\left[\int k^2 dk\ {a\Psi \Pi\sigma}\right]^\pm  
+ \left[\int k^2 dk H_{T,\tau}^{(2)} (\Pi\sigma^2)'\right]^\pm   -
\left({a_\tau \over a}\right)_- \left[\int k^2 dk H_T^{(2)} 
(\Pi\sigma^2)'\right]^\pm \right\}.
\end{equation}

\subsubsection{Even-parity perturbations}
Here we consider scalar, vector and tensor perturbations together, and
from the condition $[(\delta q_{MN})_{\rm (scalar)} + (\delta q_{MN}
)_{\rm (vector)} + 
(\delta q_{MN})_{\rm (tensor)}]^\pm = 0$, we obtain the following four
independent equations:
\begin{eqnarray}
  \label{eq:c25}
&&\Bigg[ F^2 \int k^2 dk \Big\{\zeta_0 \Pi +2v \zeta_L \Pi' + v^2[\zeta_{LL} 
\Pi + 2\phi_L (\Pi'' +{1 \over 3}k^2 \Pi)]
-2 v {l(l+1) \Pi \over \sigma} [-a^2\Psi +(a^3/\gamma)(\phi_T/a^2)_{,\tau}\cr 
&&-v\phi_T\left({\Pi' \over \Pi}-{\sigma' \over \sigma}\right)] 
+ v^2 H_T^{(2)}
{L \Pi \over \sigma^2} \Big\} \Bigg]^\pm = 0, \cr
&&\Bigg[ F\int k^2 dk \Big\{\zeta_L\Pi + 2v\phi_L[\Pi' - 
{\sigma' \over \sigma}\Pi] +[-a^2\Psi +(a^3/\gamma)(\phi_T/a^2)_{,\tau} ]
(\Pi\sigma)' + v\phi_T[(\Pi\sigma)'' -2{\sigma' \over \sigma}(\Pi\sigma)' \cr 
&&+l(l+1)(\Pi/\sigma)]
+ v H_T^{(2)} (l-1)(l+2)(\Pi\sigma)'/\sigma \Big\} \Bigg]^\pm = 0, \cr
&&\left[ \int k^2 dk \Big\{ \phi_L\Pi +2\phi_T(\Pi\sigma)' +2{\cal G}_l 
H_T^{(2)} \Big\} \right]^\pm = 0, \cr
&&\left[ \int k^2 dk \Big\{\zeta_{LL} \Pi + 2\phi_L\left[\Pi'{\sigma' 
\over \sigma} +\left({k^2 \over 3} -{l(l+1) \over 2 \sigma^2}\right)
\Pi\right] - \phi_T {l(l+1) \over \sigma^2} [(\Pi\sigma)' - 2\sigma'\Pi] - {L \over 2} H_T^{(2)}\Pi \Big\} \right]^\pm = 0,
\end{eqnarray}
where the first and second equations come from the $(00)$ and $(0a)$ 
components of the condition and the last two equations come from $(ab)$
component. In these equations the metric perturbations $\Phi_A, \Phi_H, 
\Psi$ and $H_T^{(2)}$ are given as functions of $\tau$ and $k$ at both 
regions. 

Now in the same way as in the odd-parity case, we impose a localization
 condition for  $\phi_0, \phi_L$ and $\phi_T$ as follows, using the 
function $d_\pm (\chi)$ \ (defined in Eqs.(\ref{eq:c19b}) and 
(\ref{eq:c19c}),
 so that the original local property of the 
displacement may be recovered:
\begin{eqnarray}
  \label{eq:c25a}
\int \phi_{0\pm}\Pi_\pm k^2 dk &=& \bar{\phi}_{0\pm} (\tau) 
d_\pm (\chi),\cr
\int \phi_{L\pm}\Pi_\pm k^2 dk &=& \bar{\phi}_{L\pm} (\tau) 
d_\pm (\chi),\cr
\int \phi_{T\pm}(\Pi_\pm \sigma_\pm)' k^2 dk &=& \bar{\phi}_{T\pm} (\tau) 
d_\pm (\chi),
\end{eqnarray}
where $\bar{\phi}_{0\pm} (\tau), \bar{\phi}_{L\pm} (\tau)$ and 
$\bar{\phi}_{T\pm} (\tau)$ are functions of only $\tau$. 
This condition means similarly that the total displacements have the
nonzero values only in the neighborhood of the shell. Inversely
$\phi_{0\pm}$ and $\phi_{L\pm}$ are expressed as
\begin{eqnarray}
  \label{eq:c25b}
\phi_{0\pm} &=& \bar{\phi}_{0\pm}(\tau) {1 \over k^2} \int^\infty_0 \Pi_\pm
\sigma_\pm^2  d_\pm (\chi) d\chi \simeq  \bar{\phi}_{0\pm}(\tau) \epsilon 
k^{-2} \Pi_\pm (k,\chi_b) \sigma_\pm^2 (k,\chi_b) , \cr
\phi_{L\pm} &=& \bar{\phi}_{L\pm}(\tau) {1 \over k^2} \int^\infty_0 \Pi_\pm
\sigma_\pm^2  d_\pm (\chi) d\chi \simeq  \bar{\phi}_{L\pm}(\tau) \epsilon 
k^{-2} \Pi_\pm (k,\chi_b) \sigma_\pm^2 (k,\chi_b).
\end{eqnarray}
Under this condition we find that for the above four equations there are
six unknown variables $\bar{\phi}_{0\pm}, \bar{\phi}_{L\pm}$ and 
$\bar{\phi}_{T\pm}$, representing the displacement of the shell.
But they are redundant, because only two kinds of displacements are allowed
in the even-parity case of the 2-dimensional space, as in Mukohyama's 
treatment. Here $\delta Z_a$ from
4-dimensional scalar and vector perturbations give $\phi_L \Pi Y_{lm,a}$ 
and $\phi_T (\Pi\sigma)' Y_{lm,a}$, respectively, in the similar manner.
 Accordingly we impose an additional condition to relate $\phi_{L\pm}$
and $\phi_{T\pm}$ respectively in the form of 
\begin{equation}
  \label{eq:c26}
\int \phi_{T \pm}(\Pi_\pm \sigma_\pm)' k^2 dk = 
\int \phi_{L \pm}\Pi_\pm k^2 dk \quad {\rm or} \quad
\bar{\phi}_{T\pm}(\tau) = \bar{\phi}_{L\pm}(\tau).     
\end{equation}
Then we can eliminate $\bar{\phi}_{T\pm}$ using Eq.(\ref{eq:c26}) and 
determine 
the remaining four variables solving the above four equations as follows.

Here let us assume $v = 0$ for simplicity. Then the above equations lead to
\begin{eqnarray}
  \label{eq:c27}
&&\left[- \int dk k^2 \Phi_A \Pi +(\bar{\phi}_0/a)_{,\tau} 
\right]^\pm = 0, \cr
&&\left[2(\bar{\phi}_{L,\tau} - 2{a_{,\tau} \over a}
\bar{\phi}_L) +{1 \over a} \bar{\phi}_0 - a \int dk k^2 \Psi (\Pi\sigma)'
 \right]^\pm = 0, \cr
&&\left[3\bar{\phi}_L + 2 \int k^2 dk H_T^{(2)}{\cal G}_l \right]^\pm =0, \cr
&&\left[\int k^2 dk [2 a^2 \Phi_H -{L \over 2} H_T^{(2)}] \Pi
-2a_{,\tau}\bar{\phi}_0 + 2\bar{\phi}_L {\cal{M}} \right]^\pm = 0,
\end{eqnarray}
where
\begin{equation}
\label{eq:c27a}
{\cal{M}} \equiv \int dk \int d \tilde{\chi} d(\tilde{\chi}) \Pi(\tilde{\chi})
\sigma^2 (\tilde{\chi}) {\Pi'(\chi)\over \sigma (\chi)} \left[\sigma'(\chi) 
 - {l(l+1) \Pi (\chi) \over (\Pi(\chi) \sigma (\chi))'}\right].
\end{equation}
The obtained solutions for $\bar{\phi}_{L\pm}$ and $\bar{\phi}_{0\pm}$ 
are expressed as 
\begin{eqnarray}
  \label{eq:c28} 
\bar{\phi}_{L-} &=& {2 \over 3}\int k^2 dk [H_T^{(2)}
{\cal G}_l]^\pm + \bar{\phi}_{L+}, \cr
\bar{\phi}_{L+} &=& \left\{{\cal{A}}- \int k^2 dk \left[{1 \over 4}
a\Psi(\Pi\sigma)' + {1 \over 3} H_{T,\tau}^{(2)}{\cal G}_l \right]^\pm 
+ {2 \over 3}\left({a_{,\tau} \over a}\right)_-
\int k^2 dk [H_T^{(2)}{\cal G}_l]^\pm \right\}/\left[{a_{,\tau} \over a}
\right]^\pm, 
\end{eqnarray}
and
\begin{eqnarray}
  \label{eq:c29} 
\bar{\phi}_{0+} &=& {2 a_+ a_- \over (a_+^2 -a_-^2)_{,\tau}} [{\cal{B}}/a_- -
4 {\cal{A}} a_{-,\tau}], \cr
\bar{\phi}_{0-} &=& {2 a_+ a_- \over (a_+^2 -a_-^2)_{,\tau}} [{\cal{B}}/a_+ -
4 {\cal{A}} a_{+,\tau}], 
\end{eqnarray}
where 
\begin{eqnarray}
  \label{eq:c30}
{\cal{A}} &\equiv& {1 \over 4} \int d \tau \int k^2 dk [\Phi_A \Pi]^\pm, \cr
{\cal{B}} &\equiv& \int k^2 dk \left[a^2 \Phi_H \Pi - {L \over 4}H_T^{(2)}\Pi
\right]^\pm  +\bar{\phi}_{L+} {\cal{M}}_+ - \left\{{2 \over 3}\int k^2 dk 
[H_T^{(2)}{\cal G}_l]^\pm + \bar{\phi}_{L+} \right\} {\cal{M}}_-,
\end{eqnarray}
where the definition of $\cal{M}$ is shown in Eq.(\ref{eq:c27a}) and 
$\cal{M}_\pm$ are estimated in Appendix D. 
Thus we can determine the displacements of the shell when the metric 
perturbations in both regions are given. 

As above-mentioned, we can derive
the energy-momentum tensor in the shell by substituting the obtained 
$\bar{\phi}_0, \bar{\phi}_L$ and $\bar{\phi}_T$ to  Eq.(\ref{eq:c16}), 
in which $\phi_0, \phi_L$ and $\phi_T$ can be replaced by $\bar{\phi}_0, 
\bar{\phi}_L$ and $\bar{\phi}_T$, and we find that the results are
independent of the value of $\epsilon$, as long as $\epsilon/\chi_b << 1$.

\section{Concluding remarks}

We have studied the junction conditions imposed on the metric perturbations
in the shell boundary in a four-dimensional model 
with two homogeneous regions V$^\pm$.  
As the result we found that the perturbations with different parities 
can be separated and treated independently, and that due to the 
inhomogeneity in the shell the mode coupling is caused among 
perturbations with various types and different wave-numbers in each 
parity. The displacements of the shell and the perturbed energy 
and momentum density for given metric perturbations in both regions 
can however be consistently derived.

Our results do not mean that arbitrary perturbations in both regions 
can always be adjusted, because it depends on the physical situation of 
the shell
whether the obtained energy-momentum tensor in the shell is allowed
or not. For instance, if we consider an expanding or collapsing star with
the empty external region, any energy and momentum are not expected to be
stored in the boundary and only gravitational radiation is released 
in the external region, because there are no matter perturbations.   
 
In the present study, though we treated the displacement of the 
boundary shell as spatially 3-dimensional quantities, we could recover
the original local property of the shell displacement (that it has the 
values only in the neighborhood of the shell) by imposing the localization
condition and could solve the equations for the displacements. 
 
Our result will be useful to the analyses of CMB anisotropy observed
in the nonlinear structures such as overdense regions (such as 
superclusters) and underdense regions (such as voids) in which we may 
live.
\acknowledgments{}
The author would like to thank S. Mukohyama and N. Sakai for valuable 
discussions and helpful comments.

\appendix
\section{Perturbations in a homogeneous model}

We show here the expressions for metric perturbations and harmonic functions 
in a homogeneous model with the line-element in the text, paying attention 
to the 2-dimensional parity operation $P: \theta \rightarrow \pi - \theta, 
\ \varphi \rightarrow \pi + \varphi$, where $x^2 = \theta$ and $x^3 = 
\varphi$.\cite{gsharm,rw}

\subsection{Scalar perturbations}
Metric perturbations are expressed as
\begin{eqnarray}
  \label{eq:a1}
g_{00} &=& -a^2 [1 + 2 A(\eta) Q], \cr
g_{0i} &=& -a^2 B(\eta) Q_i, \cr
g_{ij} &=& a^2 \{[1 + 2H_L(\eta) Q]\gamma_{ij} + 2H_T(\eta) Q_{ij}\}.
\end{eqnarray}
Here $Q$ is a scalar harmonic function with the wave-number $k$ satisfying
the equation
\begin{equation}
  \label{eq:a2}
(\Delta + k^2) Q = 0,
\end{equation}
where the Laplacian $\Delta$ is defined by $\Delta \phi \equiv \gamma^{ij} 
\phi_{|i} \phi_{|j}$ and $|i$ denotes the three-dimensional covariant 
derivative in the space
$dl^2 = \gamma_{ij} dx^i dx^i$. $Q_i$ and $Q_{ij}$ are defined by
\begin{eqnarray}
  \label{eq:a3}
Q_i &=& - k^{-1} Q_{|i}, \cr
Q_{ij} &=& k^{-2} Q_{|ij} + {1 \over 3} \gamma_{ij} Q.
\end{eqnarray}
Harmonic function $Q$ is expanded using the usual spherical harmonics 
$Y_{lm}$ as 
\begin{equation}
  \label{eq:a4}
Q = \Pi^n_l (\chi) Y_{lm}(\theta, \varphi),
\end{equation}
where $k^2 = n^2 -K$ for the spatial curvature $K = 1, 0, -1$.  
Then from Eq.(\ref{eq:a2}) we obtain
\begin{equation}
  \label{eq:a5}
\Pi'' + 2 (\sigma'/\sigma) \Pi' + [n^2 - l(l+1)/\sigma^2 ] \Pi = 0,
\end{equation}
where we omitted siffices of $\Pi^n_l$ for simplicity, a prime 
denotes the partial derivative with respect to $x^1 = \chi$
and $\sigma (\chi) = \sin \chi, \chi, \sinh \chi$ for $K = 1, 0, -1$, 
respectively. The normalized expressions of $\Pi$ are found in Wilson's 
paper\cite{wilson}. 
For the above-mentioned parity operation, we have 
\begin{equation}
  \label{eq:a5a}
P(Y_{lm}) = (-1)^l Y_{lm}.
\end{equation}
This property is called the even parity.

\subsection{Vector perturbations}
Metric perturbations are expressed as
\begin{eqnarray}
  \label{eq:a6}
g_{00} &=& -a^2, \cr
g_{0i} &=& -a^2 B^{(1)}(\eta) V_i, \cr
g_{ij} &=& a^2 [\gamma_{ij} + 2H^{(1)}_T(\eta) V_{ij}].
\end{eqnarray}
Here $V_i$ is a vector harmonic function with the wave-number $k$ satisfying
the equations
\begin{equation}
  \label{eq:a7}
(\Delta + k^2) V_i = 0, \quad V^i_{|i} = 0,
\end{equation}
and $V_{ij}$ is defined by
\begin{equation}
  \label{eq:a8}
V_{ij} \equiv - (2k)^{-1} [V_{i|j} + V_{j|i}]. 
\end{equation}
In vector perturbations there are two cases with different parities.

\subsubsection{The case of odd parity}

The components of vector harmonics $V_i$ are expressed as
\begin{eqnarray}
  \label{eq:a9}
V_1 &=& 0, \cr
V_2 &=& - \sigma \ \Pi Y_{lm,3}/\sin \theta,\cr
V_3 &=& \sigma \ \Pi Y_{lm,2} \sin \theta.
\end{eqnarray}
In this case we have $P(V_i) = (-1)^{l-1} V_i$ and the property is called 
the odd parity.

\subsubsection{The case of even parity}
The vector harmonics $V_i$ in another case is
\begin{eqnarray}
  \label{eq:a10}
V_1 &=& {l(l+1) \over \sigma } \Pi Y_{lm},\cr
(V_2, V_3) &=&  (\Pi \sigma)_{,1} (Y_{lm,2}, Y_{lm,3}).
\end{eqnarray}
In this harmonics, we have $P(V_i) = (-1)^l V_i$, so that the property
is called the even parity. 

\subsection{Tensor perturbations}

Tensor harmonics $G_{ij}$  satisfies
\begin{equation}
  \label{eq:a11}
{G_{ik}}^{|j}_{|j} = - k^2 G_{ik}, \ {G_i^k}_{|k} = 0, \ G_i^i = 0,
\end{equation}   
and the components are expressed as follows using $\Pi^n_l$ and $Y_{lm}$.
The tensor perturbations also have two cases with different parities.

\subsubsection{The case of odd parity}

\begin{eqnarray}
  \label{eq:a12}
G_{11} &=& 0,\cr
G_{22} &=& (\Pi \sigma^2)_{,1} (-X_{lm}/\sin \theta),\cr
G_{33} &=& (\Pi \sigma^2)_{,1} X_{lm} \sin \theta,\cr
G_{23} &=& (\Pi \sigma^2)_{,1} Z_{lm} \sin \theta,\cr
G_{12} &=&  (l-1)(l+2)\Pi (-Y_{lm,3} / \sin \theta),\cr
G_{13} &=&  (l-1)(l+2)\Pi Y_{lm,2}  \sin \theta,
\end{eqnarray}
where 
\begin{equation}
  \label{eq:a13}
X_{lm} \equiv 2(Y_{lm,23} - \cot \theta Y_{lm,3}),
\end{equation}
\begin{equation}
  \label{eq:a14}
Z_{lm} \equiv Y_{lm,22} - \cot \theta Y_{lm,2} - Y_{lm,33}/\sin^2 \theta.
\end{equation}
If we define a two-dimensional vector $W_{a}$ with the components
$W_{2} = -Y_{lm,3}/\sin \theta$ and $W_{3} = Y_{lm,2} \sin \theta$, 
it is found that $G_{ab}$ and $G_{1a}$ is reduced to
\begin{eqnarray}
  \label{eq:a14a}
G_{ab} &=& (\sigma^2 \Pi)_{,1} (W_{a\| b} + W_{b\| a}) ,\cr
G_{1a} &=&  (l-1)(l+2)\Pi W_{a}.
\end{eqnarray}
For the parity operation, we have the odd-parity property $P(G_{ij}) 
= (-1)^{l-1} G_{ij}$.  

\subsubsection{The case of even parity}

The tensor harmonics in another case are
\begin{eqnarray}
  \label{eq:a15}
G_{11} &=& {L \over \sigma^2} \Pi Y_{lm},\cr
(G_{22}, G_{33}) &=& - {L \over 2} \Pi Y_{lm}(1, \sin^2 \theta)
 + {\cal G}^n_l Z_{lm}(1, -\sin^2 \theta),\cr
G_{23} &=& {\cal G}^n_l X_{lm},\cr
G_{12} &=& (l-1)(l+2) {1 \over \sigma}(\Pi \sigma)_{,1} Y_{lm,2},\cr
G_{13} &=& (l-1)(l+2) {1 \over \sigma}(\Pi \sigma)_{,1} Y_{lm,3},
\end{eqnarray}
where $L \equiv l(l+1)(l-1)(l+2),$ and
\begin{equation}
  \label{eq:a16}
{\cal G}^n_l \equiv \sigma \sigma' \Pi_{,1} + [{1 \over 2}(l^2+l+2)
-(k^2 -2)\sigma^2] \Pi.
\end{equation}
In this case we have the even-parity property $P(G_{ij}) = (-1)^l
G_{ij}$. 

\section{$J_{\lambda \mu \nu}$}

The resulting expressions for $J_{\lambda \mu \nu}$ for three types of metric 
perturbations are shown here.

\subsection{Scalar perturbations}
\begin{eqnarray}
  \label{eq:bp1}
J_{000} &=& 2\{-a^2 \dot{\Phi}_A + \ddot{\phi}_0 - 3{\dot{a} \over a}
\dot{\phi}_0 + [3({\dot{a} \over a})^2 - {\ddot{a} \over a}]\phi_0\}Q,\cr
J_{i00} &=& 2[a^2 \Phi_A + \ddot{\phi}_L - 3{\dot{a} \over a}
\dot{\phi}_L + [2({\dot{a} \over a})^2 - {\ddot{a} \over a}]\phi_L\}Q_{,i},
 \quad (i = 1,2,3) \cr
J_{101} &=& 2\{a^2 \dot{\Phi}_H -[{\dot{a} \over a}\dot{\phi}_0 + 
 ({\ddot{a} \over a} - 3({\dot{a} \over a})^2)\phi_0]\} Q
 +2(\dot{\phi}_L -4{\dot{a} \over a}\phi_L) Q_{,11}, \cr
J_{10b} &=& 2(\dot{\phi}_L - 2 {\dot{a} \over a}\phi_L)[Q_{,1b} 
   - {\sigma' \over \sigma}Q_{,b}], \quad (b = 2,3) \cr
J_{111} &=& [2a^2 \Phi_H - 4{\dot{a} \over a}\phi_0 
-2{\dot{a} \over a}(\dot{\phi}_L - 2{\dot{a} \over a} \phi_L)] Q_{,1}
+2 \phi_L Q_{,111},\cr
J_{11b} &=& (2a^2 \Phi_H  - 2{\dot{a} \over a}\phi_0)Q_{,b}
 +2\phi_L [Q_{,11} - 2{\sigma' \over \sigma}Q_{,1} +2({\sigma' \over 
\sigma})^2 Q]_{,b}, \quad (b = 2,3)\cr
J_{122} &=& - 2 (\Phi_H a^2 + {\dot{a} \over a} \dot{\phi}_L
)\sigma^2 Q_{,1} +2\phi_L\{Q_{,122} - 2{\sigma' \over \sigma}Q_{,22} +
\sigma\sigma' Q_{,11} -[(\sigma\sigma')' - 2({\dot{a} \over a})^2 \sigma^2]
Q_{,1}\},\cr
J_{133} &=& - 2 (\Phi_H a^2 + {\dot{a} \over a} \dot{\phi}_L 
)\sigma^2 \sin^2 \theta Q_{,1} 
+2\phi_L\{Q_{,133} - 2{\sigma' \over \sigma}Q_{,33} + \sin^2 \theta 
\sigma\sigma' Q_{,11} \cr
&-& \sin^2 \theta [(\sigma\sigma')' - 
2({\dot{a} \over a})^2 \sigma^2]Q_{,1} + \sin \theta \cos \theta 
(Q_{,1}-{\sigma' \over \sigma}Q)_{,2} \},\cr
J_{123} &=& -2\phi_L \sigma^2 \sin \theta [Q/(\sigma^2 \sin \theta)
]_{,123}, \cr
J_{011} &=& [-2(a^2 \Phi_H)^. +4a\dot{a}\Phi_A - 2{\dot{a} \over a}\dot{\phi}_0]Q
 +2\phi_0\{Q_{,11} +[{\ddot{a} \over a}+({\dot{a} \over a})^2]Q\} 
-4{\dot{a} \over a}\phi_L Q_{,11}, \cr
J_{022} &=& [-2(a^2 \Phi_H)^. +4a\dot{a}\Phi_A - 2{\dot{a} \over a}\dot{\phi}_0]
\sigma^2 Q +2\phi_0 \{Q_{,22} +[{\ddot{a} \over a}+({\dot{a} \over a})^2]
\sigma^2 Q +\sigma\sigma' Q_{,1}\} \cr
&-& 4{\dot{a} \over a}\phi_L [Q_{,22} +\sigma\sigma' Q_{,1}], \cr
J_{033} &=& [-2(a^2 \Phi_H)^. +4a\dot{a}\Phi_A - 2{\dot{a} \over a}\dot{\phi}_0]
\sigma^2 \sin^2 \theta Q +2\phi_0\{Q_{,33} +[{\ddot{a} \over a}+({\dot{a} 
\over a})^2]\sigma^2 \sin^2 \theta Q \cr
&+& \sigma\sigma' \sin^2 \theta Q_{,1}
+\sin \theta \cos \theta Q_{,2}\} 
-4{\dot{a} \over a}\phi_L [Q_{,33}+\sigma\sigma' \sin^2 \theta Q_{,1}
+\sin \theta \cos \theta Q_{,2}], \cr
J_{023} &=& 2(\phi_0 - 2{\dot{a} \over a}\phi_L)(Q_{,23} - \cot \theta 
Q_{,3}), \cr
J_{01b} &=& 2(\phi_0 - 2{\dot{a} \over a}\phi_L)(Q_{,1b} - {\sigma' \over 
\sigma}
Q_{,b}), \quad (b = 2, 3) \cr
J_{00i} &=& [-2a^2 \Phi_A + 2(\dot{\phi}_0 -2{\dot{a} \over a}\phi_0) 
-2{\dot{a} \over a}(\dot{\phi}_L -2{\dot{a} \over a}\phi_L)] Q_{,i}, \quad 
(i= 1,2,3). 
\end{eqnarray}

\subsection{Vector perturbations}
\begin{eqnarray}
  \label{eq:bp2}
J_{000} &=& 0, \cr
J_{100} &=& 2[\ddot{\phi}_T - 6{\dot{a} \over a}\dot{\phi}_T +2(-{\ddot{a} 
\over a}+2({\dot{a} \over a})^2)\phi_T -a^2(\dot{\Psi} +{\dot{a} 
\over a}\Psi)] V_1, \cr
J_{101} &=& 2(\dot{\phi}_T - 2{\dot{a} \over a}\phi_T) V_{1,1}, \cr
J_{10b} &=& a^2 \Psi (V_{b,1} - V_{1,b}) +2(\dot{\phi}_T - 2{\dot{a} \over 
a}\phi_T) (V_{1,b}-{\sigma' \over \sigma}V_b), \quad (b = 2, 3) \cr 
J_{111} &=& [2a \dot{a}\Psi - 2{\dot{a} \over a}(\dot{\phi}_T-
2{\dot{a} \over a} \phi_T)] V_1 + 2\phi_T V_{1,11}, \cr
J_{122} &=& 2{\dot{a} \over a}\sigma^2 (a^2\Psi -\dot{\phi}_T)V_1
 +2\phi_T[V_{1,22} +\sigma\sigma'V_{1,1} - 2{\sigma' \over \sigma}V_{2,2}
 -(\sigma\sigma')'V_1 +2({\dot{a} \over a})^2\sigma^2V_1], \cr
J_{133} &=& 2{\dot{a} \over a}\sigma^2 \sin^2 \theta (a^2\Psi 
-\dot{\phi}_T)V_1 +2\phi_T[V_{1,33} +\sigma\sigma'\sin^2 \theta V_{1,1} 
- 2{\sigma' \over \sigma}V_{3,3} \cr
&-&(\sigma\sigma')'\sin^2 \theta V_1 
+2({\dot{a} \over a})^2\sigma^2 \sin^2 \theta V_1 
+\sin \theta \cos \theta (V_{1,2} -2{\sigma' \over \sigma}V_2)], \cr
J_{123} &=& 2\phi_T[V_{1,23} - {\sigma' \over \sigma}(V_{2,3}+V_{3,2})
+\cot \theta (-V_{1,3} + 2{\sigma' \over \sigma}V_3)], \cr  
J_{11b} &=& 2\phi_T[V_{1,1a} - {\sigma' \over \sigma}(V_{b,1}+V_{1,b})
 + 2({\sigma' \over \sigma})^2 V_b], \quad (b = 2,3) \cr   
J_{00i} &=& [2a\dot{a}\Psi - 2{\dot{a} \over a}(\dot{\phi}_T - 
 2{\dot{a} \over a}\phi_T)] V_i, \quad (i = 1,2,3) \cr
J_{011} &=& -2(a^2 \Psi + 2{\dot{a} \over a}\phi_T) V_{1,1}, \cr
J_{022} &=& -2(a^2 \Psi + 2{\dot{a} \over a}\phi_T) (V_{2,2} +
 \sigma\sigma' V_1), \cr
J_{033} &=& -2(a^2 \Psi + 2{\dot{a} \over a}\phi_T) (V_{3,3} +
 \sigma\sigma' \sin^2 \theta V_1), \cr
J_{023} &=& -(a^2 \Psi + 2{\dot{a} \over a}\phi_T) (V_{2,3} +V_{3,2}
 - 2\cot \theta V_3), \cr
J_{01b} &=& -(a^2 \Psi + 2{\dot{a} \over a}\phi_T) (V_{1,b} +V_{b,1}
 - 2 {\sigma' \over \sigma} V_b). \quad (b = 2,3) 
\end{eqnarray}

\subsection{Tensor perturbations}
\begin{eqnarray}
  \label{eq:bp3}
J_{000} &=& J_{001} = J_{002} = J_{003} = J_{100} = 0, \cr
J_{i0j} &=& -2(a^2 \dot{H}_T^{(2)}) G_{ij}, \quad (i,j = 1,2,3) \cr
J_{111} &=& 2a^2 \dot{H}_T^{(2)} G_{11,1}, \cr
J_{122} &=& 2a^2 \dot{H}_T^{(2)}[2G_{12,2} - G_{22,1} + 
2 \sigma\sigma' G_{11}], \cr
J_{133} &=& 2a^2 \dot{H}_T^{(2)}[2G_{13,3} - G_{33,1} + 
2 \sigma\sigma'\sin^2 \theta G_{11} +2\sin \theta \cos \theta G_{12}], \cr
J_{123} &=& 2a^2 \dot{H}_T^{(2)}[G_{12,3}+ G_{31,2} -G_{23,1}- 2\cot \theta
G_{13}], \cr
J_{11b} &=& 2a^2 \dot{H}_T^{(2)}[G_{11,b} - 2{\sigma' \over \sigma}G_{1b}].
\quad (b = 2,3)
\end{eqnarray}
%

\section{The intrinsic quantity $\Theta_{MN}$ in the shell}
The expressions of $\Theta_{MN}$ are shown for three types of metric 
perturbations, where $M$ and $N$ take the values $0, 2$ and $3$. 

\subsection{Scalar perturbations}
\begin{eqnarray}
  \label{eq:cp1}
\Theta_{00} &=& {2 \over a^3}\left\{{\sigma' \over \sigma}a^2(\Phi_H +2\Phi_A)Q
 +a^2 \Phi_H Q_{,1}\right\} +{2 \over a^3}{\sigma' \over \sigma}(-2\dot{\phi}_0
+{\dot{a} \over a}\phi_0)Q \cr
&+& {2 \over a} {\dot{a} \over a}\dot{\phi}_L Q_{,1}
-{1 \over a^3}\phi_L \{{1\over \sigma^2} [(Q_{,1} -2{\sigma' \over \sigma}
Q)_{,22} +(Q_{,1} -2{\sigma' \over \sigma}Q)_{,33}/\sin^2 \theta]\cr
&-& 2[{\sigma'' \over \sigma} +({\sigma' \over \sigma})^2 -2({\dot{a} 
\over a})^2] Q_{,1} 
+ {1\over \sigma^2}\cot \theta (Q_{,1}-2{\sigma' \over \sigma} Q)_{,2}\}\cr
&+& {2v \over a^3} a^2[\dot{\Phi}_H +{\dot{a} \over a}(3\Phi_H+ 
\Phi_A)]Q + {2 \dot{v} \over a} a^2 \Phi_A Q \cr
&-& {v \over a^3}\phi_0 \left[\left(2 {\ddot{a} \over a} +4({\dot{a} 
\over a})^2 - {l(l+1) \over \sigma^2}\right)Q + 
 4{\sigma' \over \sigma}Q_{,1}\right] 
- {2 \over a^3}(3 {\dot{a} \over a}v + \dot{v})(\dot{\phi}_0 - 
{\dot{a} \over a}\phi_0)Q \cr
&+& {2 \over a^2}{\dot{a} \over a}\dot{\phi}_0 Q 
+ {2v \over a^3}\phi_L {\dot{a} \over a}[Q_{,11} 
+ {1 \over \sigma^2}(Q_{,22} +Q_{,33}/\sin^2 \theta) \cr
&+& 4{\sigma' \over \sigma}Q_{,1} +
{1\over \sigma^2}\cot \theta Q_{,2}] -{2v \over a^3}{\sigma' \over \sigma} 
\dot{\phi}_L Q_{,1} + 0(v^2), \cr
\Theta_{22} &=& \Theta_s +
 {1 \over a}\phi_L \{(Q_{,133} -2{\sigma' \over \sigma}Q_{,33})/
 \sin^2 \theta  + \cot \theta (Q_{,12} - 2{\sigma' \over 
\sigma} Q_{,2}) - 4{\sigma' \over \sigma} Q_{,22} \} + 0(v), \cr
\Theta_{33} &=& \Theta_s \sin^2 \theta +
 {1 \over a}\phi_L \{(Q_{,122} -2{\sigma' \over \sigma}Q_{,22})
\sin^2 \theta - 4{\sigma' \over \sigma}(Q_{,33} +\sin \theta \cos \theta 
Q_{,2}) \} + 0(v),\cr
\Theta_{0b} &=& -{1 \over a^2}(\dot{\phi}_L - 2{\dot{a} \over a} \phi_L)
(Q_{,1} + {\sigma' \over \sigma}Q)_{,b} - {2 \over a^2}{\sigma' \over 
\sigma}\phi_0 Q_{,b} -{2 \over a^2}\dot{v}(\dot{\phi}_L -2{\dot{a} 
\over a}\phi_L + \phi_0) Q_{,b} \cr
&-& {v \over a^2} \{[2a^2 (\Phi_H -\Phi_A) +2\dot{\phi}_0 +4{\dot{a} 
\over a}(\dot{\phi}_L
-2{\dot{a} \over a}\phi_L)]Q_{,b} +2\phi_L [Q_{,11} +2{\sigma' \over \sigma}
Q_{,1} \cr
&-& 2({\sigma' \over \sigma})^2 Q]_{,b}\} + 0(v^2), \quad (b = 2,3)
\cr
\Theta_{23} &=& {1 \over a}\{\phi_L [Q_{,12} -2{\sigma' \over \sigma}Q_{,2}
-\cot \theta  (Q_{,1} - 2{\sigma' \over \sigma}Q)]_{,3} 
-v (\phi_0 - 2{\dot{a} \over a}\phi_L)(Q_{,2} - \cot \theta Q)_{,3} \cr
&+& 2(2{\sigma' \over \sigma} -3{\dot{a} \over a}v - \dot{v}) \phi_L
(Q_{,23} - \cot \theta Q_{,3})\} + 0(v^2) ,  
\end{eqnarray}
where
\begin{eqnarray}
\Theta_s /\sigma^2 &\equiv& -{1 \over a}a^2(\Phi_A+ \Phi_H)Q_{.1}
-5{\sigma' \over \sigma}{1 \over a}(a^2 \Phi_H - {\dot{a} \over a}\phi_0)Q
- {1 \over 2a}(\ddot{\phi}_L -2{\dot{a} \over a}\dot{\phi}_L) Q_{,1} \cr
&+& {1 \over a}\phi_L \{ - {\sigma' \over \sigma}Q_{,11} +
[{\sigma'' \over \sigma} -3({\sigma' \over \sigma})^2]Q_{,1} + 
[{\ddot{a} \over a} -({\dot{a} \over a})^2]Q_{,1}  \} + 0(v).
\end{eqnarray}
\subsection{Vector perturbations}
\begin{eqnarray}
  \label{eq:cp2}
\Theta_{00} &=& -{1 \over a^3}\Bigl\{{\dot{a} \over a}(a^2\Psi -
\dot{\phi}_T)V_1 +\phi_T [-2 {\sigma' \over \sigma}V_{1,1} +
4({\dot{a} \over a})^2 V_1 + {1 \over \sigma^2} \Bigl(V_{1,22} - 
2{\sigma' \over \sigma} V_{2,2} \cr
&+&(V_{1,33} - 2{\sigma' \over \sigma}V_{3,3})/\sin^2 \theta + 
\cot \theta (V_{1,2} -2{\sigma' \over \sigma}V_2) \Bigr) 
+ 2{\sigma' \over \sigma} V_{1,1} - {2(\sigma\sigma')' \over \sigma^2}
V_1]\Bigr\} \cr
&+&{v \over a^3}\Bigl\{2{\dot{a} \over a}\phi_T V_{1,1} -2[a^2\Psi+ 
a^2(\phi_T/a^2)^.]{\sigma' \over \sigma}V_1 \cr
&+& (a^2\Psi +2{\dot{a} \over a}\phi_T)[{1 \over \sigma^2}(V_{2,2} + V_{3,3}
/\sin^2 \theta) + 2{\sigma' \over \sigma}V_1] \cr
&+& 4{\sigma' \over \sigma}[-a^2\Psi +a^2(\phi_T/a^2)^.]V_1 \Bigr\} 
+ 0(v^2), \cr
\Theta_{22} &=& \Theta_v +{ 1 \over a}\phi_T [(V_{1,33}-
2{\sigma' \over \sigma}V_{3,3})/\sin^2 \theta -4{\sigma' \over \sigma}V_{2,2}
 + \cot \theta (V_{1,2} -2{\sigma' \over \sigma}V_2)]\cr
&-& {2 \over a}\phi_T (3{\dot{a} \over a}v +  \dot{v})
V_{2,2} - {v \over a}(a^2\Psi +2{\dot{a} \over a}\phi_T)V_{3,3}
/\sin^2 \theta + 0(v^2), \cr
\Theta_{33} &=& \Theta_v \sin^2 \theta
+{ 1 \over a}\phi_T [(V_{1,22}-
2{\sigma' \over \sigma}V_{2,2})\sin^2 \theta - 4{\sigma' \over \sigma}
(V_{3,3} + \sin \theta \cos \theta V_2) ] 
- {2 \over a}\phi_T (3{\dot{a} \over a}v +  \dot{v})
(V_{3,3} \cr
&+& \sin \theta \cos \theta V_2) 
- {v \over a}(a^2\Psi +2{\dot{a} \over a}\phi_T)V_{2,2} \sin^2 \theta
 + 0(v^2)\cr 
\Theta_{23} &=& -{1 \over a}\phi_T [V_{1,23} +{\sigma' \over \sigma}
(V_{2,3} +V_{3,2}) 
-\cot \theta (V_{1,3} +2{\sigma' \over \sigma}V_3)] \cr
&+& {v \over 2a}[a^2 \Psi +2({\dot{a} \over a}-{2 \sigma' \over \sigma})
\phi_T](V_{2,3} +V_{3,2} -2\cot \theta V_3), \cr
\Theta_{0b} &=& -{1 \over a^2} \{{1 \over 2} a^2\Psi (V_{b,1}-V_{1,b})
+(\dot{\phi}_T -2{\dot{a} \over a}\phi_T)(V_{1,b} -{\sigma' \over \sigma}V_b)
\cr
&+& (2{\sigma' \over \sigma}+3{\dot{a} \over a}v + \dot{v})[-a^2\Psi
+a^2(\phi_T/a^2)^.]V_b \} \cr
&-& { v \over a^2} \{[a\dot{a}\Psi -{\dot{a} \over a}(\dot{\phi}_T
-2{\dot{a} \over a}\phi_T)]V_b + \phi_T[V_{1,1b} -{\sigma' \over \sigma} 
(V_{b,1}+V_{1,b}) +2({\sigma' \over \sigma})^2V_b]\} \cr
&+& [-{1 \over a}v\phi_T (2{\sigma' \over \sigma}+3{\dot{a} \over a}v 
+ \dot{v}) ](V_{1,b}+V_{b,1}- 2{\sigma' \over \sigma}V_b) + 0(v^2).
 \quad (b = 2,3)
\end{eqnarray}
where
\begin{eqnarray}
\Theta_v &\equiv& -{\sigma^2 \over a}{\sigma' \over \sigma}\phi_T
[V_{1,1} + 4{\sigma' \over \sigma} V_1]
-{ \sigma^2 \over a}[\ddot{\phi}_T - 6{\dot{a} \over a}
\dot{\phi}_T +2(-{\ddot{a} \over a}+2({\dot{a} \over a})^2)\phi_T \cr
&-& a^2(\dot{\Psi}+{\dot{a} \over a}\Psi)]V_1 
+{\sigma^2 \over a} \{{\dot{a} \over a}(a^2\Psi-\dot{\phi}_T)V_1
+ \phi_T[{\sigma' \over \sigma}V_{1,1} \cr
&+& 2({\dot{a} \over a})^2V_1 -
({\sigma' \over \sigma})'  \sigma^{-2} V_1] \} 
-{\sigma^2 \over a} \{[2v(\ddot{\phi}_T -{\dot{a} \over a}\phi_T)
+ \dot{v}\phi_T] V_{1,1} + {\sigma' \over \sigma}v[a^2\Psi \cr
 &+& a^2(\phi_T/a^2)^.]V_1\}
-2{\sigma^2 \over a}\phi_T (3{\dot{a} \over a}v + \dot{v})
{\sigma' \over \sigma} V_1 
- {v \over a}(a^2\Psi +2{\dot{a} \over a}\phi_T)\sigma \sigma' V_1 + 0(v^2).
\end{eqnarray}

\subsection{Tensor perturbations}
\begin{eqnarray}
  \label{eq:cp3}
\Theta_{00} &=& -a^{-1}H_T^{(2)}[\sigma^{-2}(2G_{12,2} -G_{22,1} +
(2G_{13,3} -G_{33,1})/\sin^2 \theta) +6{\sigma' \over \sigma}G_{11} \cr
&+& 2 \cot \theta \sigma^{-2}G_{12}] + 0(v), \cr
\Theta_{22} &=& a H_T^{(2)} [3{\sigma \sigma'} G_{11}
 + (2G_{13,3} -G_{33,1})/\sin^2 \theta +2\cot \theta G_{12}] + 0(v), \cr
\Theta_{33} &=& a H_T^{(2)} [3{\sigma \sigma'} G_{11}
 + (2G_{12,2} -G_{22,1})] \sin^2 \theta + 0(v), \cr
\Theta_{23} &=&  a H_T^{(2)} [4{\sigma' \over \sigma} G_{23} -(G_{12,3}
+G_{13,2} -G_{23,1} - 2\cot \theta G_{13})] + 0(v), \cr
\Theta_{0b} &=& -\dot{H}_T^{(2)} G_{1b} + 0(v). \quad (b = 2,3) 
\end{eqnarray}
%
%
\section{Integrals ${\cal{L}}$ and ${\cal{M}}$}
The integrals ${\cal{L}}$ and ${\cal{M}}$ in Eqs. (\ref{eq:c22b}) and 
(\ref{eq:c27a}) are rewritten here using the the orthonormal relations 
(\ref{eq:c19e}).

First we have
\begin{eqnarray}
  \label{eq:d1}
{\cal{L}} &=& [\sigma' (\chi)/\sigma (\chi)] \int d\tilde{\chi} d(\tilde{\chi}) 
\int dk \Pi(\tilde{\chi})
\sigma(\tilde{\chi}) \Pi(\chi) \sigma (\chi) \cr
&=& [\sigma' (\chi)/\sigma (\chi)] d(\chi).
\end{eqnarray}
Since $d(\chi) = 1$ in the neighborhood of the shell, we obtain 
\begin{equation}
  \label{eq:d2}
{\cal{L}}_\pm = (\sigma'/\sigma)_\pm.
\end{equation}

Next we notice that
\begin{equation}
  \label{eq:d3}
{\Pi\Pi' \over (\Pi\sigma)'} = {\Pi \over \sigma}\left(1 - {\Pi\sigma' \over 
(\Pi\sigma)'}\right)
\end{equation}
and the factor $|\Pi\sigma'/(\Pi\sigma)'|$ is $\ll 1$, if assume that 
the perturbations shorter enough than the shell size contribute dominantly
to the integral. Then ${\cal{M}}$ reduces to
\begin{eqnarray}
  \label{eq:d4}
{\cal{M}} &=& {\partial \over \partial \chi} \int d\tilde{\chi} 
d(\tilde{\chi})
\sigma(\tilde{\chi}) \int dk \Pi(\tilde{\chi}) \sigma(\tilde{\chi}) 
\left[\left(\Pi(\chi) \sigma (\chi) {\sigma'(\chi) \over \sigma (\chi)}
\right)' - 
{\Pi (\chi) \sigma'(\chi) \over \sigma (\chi)} \left({\sigma' (\chi) \over 
\sigma (\chi)}\right)'\right] \cr
 &-& l(l+1)  \int d\tilde{\chi} d(\tilde{\chi})\sigma (\tilde{\chi})
\int dk \Pi(\tilde{\chi}) \sigma(\tilde{\chi}) {\Pi (\chi) \sigma (\chi) 
\over \sigma^3 (\chi)} \left(1 - {\Pi (\chi)\sigma' (\chi)\over (\Pi (\chi)
\sigma (\chi))'}\right) \cr
&=& \left\{\sigma \left[({\sigma' \over \sigma^2})' - {1 \over \sigma}({\sigma' \over 
\sigma})'\right] - { l(l+1) \over \sigma^2} \left(1 - 0\langle {\Pi\sigma' \over 
(\Pi\sigma)'} \rangle \right)\right\} d(\chi) \cr
&=& - {1 \over \sigma^2} \left[(\sigma')^2 + l(l+1) \left(1 - 0\langle 
{\Pi\sigma' \over (\Pi\sigma)'} \rangle \right)\right] d(\chi),
\end{eqnarray}
where $\langle {\Pi\sigma' / (\Pi\sigma)'} \rangle$ represents the 
root-mean-square of $\Pi\sigma' / (\Pi\sigma)'$ in the $k$-space.
Accordingly, we obtain
\begin{equation}
  \label{eq:d5}
{\cal{M}}_\pm \simeq - {1 \over \sigma^2_\pm} [(\sigma')^2_\pm + l(l+1)].
\end{equation}
%




\end{document}